\documentclass{jltp}
\usepackage{amsmath}
\title{Superfluid Phases of $^3$He in Aerogel}
\author {I.A.Fomin\address{P.L.Kapitza Institute for Physical Problems,\\
Kosygina 2, 119334 Moscow, Russia}}
\date{}
\begin{document}
\maketitle{}
\begin{abstract}
Formal derivation of criterion for selection of superfluid phases of
$^3$He in aerogel is presented. At the strength of the derived criterion
variation of the order parameter of B-like phase in magnetic field differs
from that of B-phase of pure $^3$He. Possible observable consequence of this
difference is discussed.

\end{abstract}

\section {INTRODUCTION}
In pure $^3$He two superfluid phases -- A and B are observed.
Existence of more then one superfluid phase is a manifestation of
unconventional character of Cooper pairing in this liquid. In both phases
Cooper pairs have angular momentum $l=1$ and spin $s=1$. Their order parameters
are 3$\times$3 complex matrices $A_{\mu j}$, index  $\mu$ refers to spin and
$j$ to the orbital projection.  The particular form of order parameter
for the A-phase is
$$
A^{ABM}_{\mu j}=\Delta\frac{1}{\sqrt{2}}\hat d_{\mu}(\hat m_j+i\hat
n_j),                                                           \eqno(1)
$$
where
${\bf\hat m}$,  ${\bf\hat n}$ are mutually orthogonal unit vectors in orbital
space and ${\bf\hat d}$ - a unit vector in spin space, $\Delta$ is an overall
amplitude.  For B-phase the order parameter is
$$
A^{BW}_{\mu j} = \Delta\frac{1}{\sqrt{3}}e^{i\varphi}R_{\mu j}, \eqno(2)
$$
where $R_{\mu j}$ is a real orthogonal
matrix.  In a presence of aerogel the property of superfluidity survives
\cite{parpia} and two superfluid phases (called A-like and B-like) are observed
\cite{osher,halp} as well. In a simplified form aerogel can be represented
as an array of randomly distributed and randomly oriented strands of diameter
$d\approx 30$\AA. For 98\% aerogel the average distance between the strands is
$l\approx 200$\AA, i.e. comparable with the correlation length of the
condensate $\xi_0$, which varies depending on a pressure in the interval
160-500\AA. Aerogel makes condensate of Cooper pairs spatially
nonuniform. Scattering of quasiparticles on the strands has depairing effect on
components of the order parameter which extends over a distance of the order of
$\xi_0$ \cite{Rainer}.
An array of strands induces random variation of the order parameter in space.
It is convenient to represent the order parameter $A_{\mu j}({\bf r})$
as a sum of its average value and a fluctuating increment:
 $A_{\mu j}({\bf r})=\bar A_{\mu j}+a_{\mu j}({\bf r})$. By the definition
 the average $\bar a_{\mu j}({\bf r})\equiv <a_{\mu j}({\bf r})>$=0.
 Averaging is assumed over a volume with a characteristic size
 exceeding a length for which the distribution of the strands is correlated.
 Onset of the long-range order is formally expressed by the condition
 $\bar A_{\mu j}\ne$0. The form of  $\bar A_{\mu j}$ describes the change of
 symmetry at the phase transition and symmetry properties of the
 emerging phase. Identification of  $\bar A_{\mu j}$  for A-like and
 B-like phases is a key question for theoretical interpretation of their
 properties. Fluctuating part $a_{\mu j}({\bf r})$  influences averaged
 properties of the superfluid
 phases as well and because of nonlinear dependence
 of free energy of these phases on the order parameter fluctuations can
 influence a choice of  $\bar A_{\mu j}$. This influence has been discussed
 recently\cite{fom1,fom2,fom3}. In the present paper some of the previous
 discussion is summarised and new application of the developed approach is
 presented.

 \section {RANDOM FIELD}
 Since the average distance between the strands is of the order of $\xi_0$
 perturbation of the order parameter at a given point appears as effect of
 several neighbouring strands.  In a vicinity of $T_c$ correlation length
 $\xi(T)\gg\xi_0$ and effect of aerogel can be represented by a local
 random field. Its form is determined by
 symmetry argument and natural assumptions about properties
of aerogel. Most important assumption is that the scattered quasiparticles
do not exchange their spins with the strands. In practice this condition is
secured by a coverage of the strands by one-two monolayers of atoms of $^4$He.
With this precaution aerogel can influence directly only orbital part of the
order parameter.  A principal contribution of interaction with aerogel to the
free energy functional is of the second order in the order parameter and it can
be written in a form

$$
F_{\eta}=N(0)\int \eta_{jl}({\bf r})A_{\mu j}A_{\mu l}^*d^3r , \eqno(3)
$$
where $N(0)$ is the density of states on Fermi level and
$\eta_{jl}({\bf r})$ -- a random static tensor field. Further properties of
$\eta_{jl}({\bf r})$ can be found on a strength of general argument. Since
$F_{\eta}$ must be real $\eta_{jl}({\bf r})$ is Hermitian matrix and to secure
 $t\to -t$ invariance of the energy it must be symmetrical. So
 $\eta_{jl}({\bf r})$ is real and symmetrical tensor field.
Characteristic amplitude of the random field can be estimated with the use of
 results of the paper\cite{Rainer}. By the order of magnitude it is
$|\eta_{jl}|\sim(\xi_0 d/l^2)$. For  98\% aerogel this ratio is $\sim 1/10$
and effect of aerogel can be treated as a perturbation.
With the account of interaction term (3) Ginzburg-Landau (GL) functional takes
the following form:
$$
 F_{GL}=N(0)\int d^3r[\tau A_{\mu j}A_{\mu j}^*+
\eta_{jl}({\bf r})A_{\mu j}A_{\mu l}^*+\frac{1}{2}\sum_{s=1}^5 \beta_sI_s+
$$
$$
\frac{1}{2}\left(K_1\frac{\partial A_{\mu l}}{\partial x_j}
\frac{\partial A^*_{\mu l}}{\partial x_j}+
K_2\frac{\partial A_{\mu l}}{\partial x_j}\frac{\partial A^*_{\mu j}}{\partial
x_l}+K_3\frac{\partial A_{\mu j}}{\partial x_j}\frac{\partial A^*_{\mu
l}}{\partial x_l}\right)\Bigr] ,           \eqno(4)
$$
where $\tau=(T-T_c)/T_c$, $I_s$ -  4-th order invariants in the expansion of
the free energy over $A_{\mu j}$  (cf.\cite{vollh}).
Coefficients $\beta_1,...\beta_5,K_1,K_2,K_3$ are phenomenological constants.
In what follows for their evaluation the weak coupling values are assumed.
 Possible random corrections to
the gradient terms contain extra power of the wave-vector and will not influence
 the following discussion.
Comparison of two first terms in the functional indicates a physical meaning
of the tensor $\eta_{jl}({\bf r})$ -- its isotropic part
$\frac{1}{3}\eta_{mm}({\bf r})\delta_{jl}$ describes  local variations of
$T_c=T_c({\bf r})$ caused by fluctuations of the density of scatterers.
Anisotropic part
$\eta_{jl}({\bf r})-\frac{1}{3}\eta_{mm}({\bf r})\delta_{jl}
\equiv\eta^{(a)}_{jl}$ describes the local splitting of  $T_c$ for different
projections of angular momenta caused by the local breaking of spherical
symmetry by the aerogel strands.

\section {CRITERION FOR SELECTION OF THE PHASES}
The order parameters of stable phases are found as minima of the functional
(4). Because of a presence of the field $\eta_{jl}({\bf r})$ the gradient
terms can not be omitted. According to the procedure
developed for a scalar order parameter in a scalar random field
\cite{LarkOv} the order parameter
 $A_{\mu j}({\bf r})=\bar A_{\mu j}+a_{\mu j}({\bf r})$ is substituted
 in the GL equation, corresponding to the functional (4) and terms up to
 the second order in $a_{\mu j}$ and $\eta_{jl}$ are kept.
 After averaging over coordinate ${\bf r}$ one arrives at the equation

$$
\tau\bar A_{\mu
j}+\frac{1}{2}\sum_{s=1}^5 \beta_s\bigl[\frac{\partial I_s}{\partial A^*_{\mu
j}}+ \frac{1}{2}\bigl( \frac{\partial^3 I_s}{\partial A^*_{\mu j}\partial
A_{\nu n}\partial A_{\beta l}} <a_{\nu n}a_{\beta l}>+
$$
$$ 2\frac{\partial^3
I_s}{\partial A^*_{\mu j}\partial A^*_{\nu n} \partial A_{\beta l}}<a^*_{\nu
n}a_{\beta l}>\bigr)\bigr]= -<a_{\mu l}\eta_{lj}>.              \eqno(5)
$$
Along with $\bar A_{\mu j}$ the obtained equation  contains averages of
binary products of fluctuations $<a_{\nu n}a_{\beta l}>$ etc..
Equations for $a_{\mu j}$ are obtained by separation of fast varying terms
in the GL equation
$$
\tau a_{\mu
j}+\frac{1}{2}\sum_{s=1}^5\beta_s\bigl[ \frac{\partial^2 I_s}{\partial A^*_{\mu
j}\partial A_{\nu n}}a_{\nu n}+ \frac{\partial^2 I_s}{\partial A^*_{\mu
j}\partial A^*_{\nu n}}a^*_{\nu n}\bigr]-
$$
$$ \frac{1}{2}K\left(\frac{\partial^2
a_{\mu j}}{\partial x_l^2}+ 2\frac{\partial^2 a_{\mu l}}{\partial x_l \partial
x_j}\right)= -\bar A_{\mu l}\eta_{lj}.               \eqno(6)
$$

To proceed further one needs to introduce a nontrivial modification of
the procedure of Ref.\cite{LarkOv}  caused by a more complicated form
of the order parameter and its higher degeneracy.
Eq.(6) and its complex conjugate form a linear inhomogeneous system.
The corresponding homogeneous system has nontrivial solutions as a consequence
of the degeneracy of the unperturbed GL functional (functional (4) without
the random field) with respect to orbital rotations of the order parameter.
These solutions are the increments of $\bar A_{\mu j}$ and $\bar A^*_{\mu j}$
at the infinitesimal rotation $\theta_n$:

$$
\omega_{\mu j}=\theta_ne^{jnr}\bar A_{\mu r},
\omega^*_{\mu j}=\theta_ne^{jnr}\bar A^*_{\mu r},         \eqno(7)
$$
where $e^{jnr}$ is the absolutely antisymmetric tensor.
Eq. (6) is solved by Fourier transformation.
For further argument only a character of singularity of
$a_{\mu j}({\bf k})$ at $k\to 0$ is important. To find it one can neglect
anisotropy of the gradient terms in Eq. (6) and to substitute
$\frac{1}{2}\bar K\left(\frac{\partial^2 a_{\mu j}}{\partial x_l^2}\right)$
instead of
$\frac{1}{2}K\left(\frac{\partial^2 a_{\mu j}}{\partial x_l^2}+
2\frac{\partial^2 a_{\mu l}}{\partial x_l \partial x_j}\right)$.
Then  Eq.(6) is multiplied
by $\omega^*_{\mu j}$. The result is summed with its
complex conjugate and the obtained equation is solved with respect to the
projection $a^{\omega}({\bf k})\equiv a_{\mu j}({\bf k})\omega^*_{\mu j}+
a^*_{\mu j}({\bf k})\omega_{\mu j}$:

$$
a^{\omega}({\bf k})=-\frac{2}{k^2\bar K}
\left(\theta_n e^{jnr}Q_{rl}\eta^{(a)}_{lj}({\bf k})\right),     \eqno(8)
$$
where $Q_{rl}=\bar A_{\mu r}\bar A^*_{\mu l}+\bar A_{\mu l}\bar A^*_{\mu r}$ .
This projection gives a contribution to the averages
 $<a_{\nu n}a_{\beta l}>$  which is proportional to
$$
\int<a^{\omega}({\bf k})a^{\omega}(-{\bf k})>d^3k \sim
$$
$$
\theta_p e^{jpr}Q_{rl}\theta_q e^{mqs}Q_{sn}
\int<\eta^{(a)}_{lj}({\bf k})\eta^{(a)}_{mn}(-{\bf k})>\frac{d^3k}{k^4}.
                                                                 \eqno(9)
$$
Form of the correlation function in the integrand after averaging over
 direction of ${\bf k}$ is determined by the symmetry properties of
 $\eta^{(a)}_{lj}$:
$$
\int<\eta^{(a)}_{lj}({\bf k})\eta^{(a)}_{mn}(-{\bf k})>\frac{do}{4\pi}=
\Phi(k)[2\delta_{lj}\delta_{mn}-3(\delta_{lm}\delta_{jn}+
\delta_{ln}\delta_{jm})].                                         \eqno(10)
$$
Finally
$$
\int<a^{\omega}({\bf k})a^{\omega}(-{\bf k})>d^3k \sim
$$

$$
\theta_p\theta_q[e^{npr}e^{lqs}Q_{rl}Q_{sn}+(\delta_{pq}\delta_{rs}-
\delta_{ps}\delta_{qr})Q_{rn}Q_{sn}]\int\frac{\Phi(k)}{k^2}dk.   \eqno(11)
$$
The integral in the right-hand side diverges at the lower limit and
the averaged products $<a^{\omega}a^{\omega}>$ give diverging contributions
to Eq. (5). To make the iteration scheme Eqs. (5),(6) consistent
one has to require the coefficient in front of the diverging integral to be
zero.  Equating to zero the expression in square brackets in Eq. (11) one
arrives at $Q_{rn}=q\delta_{rn}$, where $q$ is a real number. In terms of the
order parameter this condition reads as:
$$
\bar A_{\nu r}\bar A^*_{\nu
j}+\bar A_{\nu j}\bar A^*_{\nu r}= const\cdot\delta_{rj}.  \eqno(12)
$$
This is a
necessary condition for existence of a superfluid phase with the order
parameter $\bar A_{\nu r}$. The phases meeting this criterion are
``quasi-isotropic" i.e. the energy of their interaction with aerogel does not
change at the arbitrary orbital rotation:
$$
\eta^{(a)}_{jl}A_{\mu j}A_{\mu l}^*=0.                            \eqno(13)
$$
Tensor of superfluid densities for these phases is isotropic and as a
consequence their order parameter is not oriented by a superflow.
It has been shown before\cite{fom3} that when condition (13)
is met the general argument of Imry and Ma\cite{imry}  does not apply and the
long-range order is not destroyed by the random field $\eta_{jl}({\bf r})$.  It
might be well to point out here the difference between the argument of Imry and
Ma and the argument presented in this section. In the former case the
order parameter is assumed to be fixed by minimization of free energy
without the random field. Possible
disordering is caused by a random walk of the order parameter over its space of
degeneracy. No other change of the order parameter  is admitted. In the latter
case all variations of the order parameter $A_{\mu j}$ are considered. That
makes possible an adjustment of the order parameter to the random field.
To emphasize a property of phases, which satisfy criterion (12) or (13) to
resist disordering action of the random field a term ``robust" was suggested
\cite{fom2}.

Now let us discuss a region of applicability of the above argument.
Dimensionless distance from $T_c$: $|\tau|=(T_c-T)/T_c$ is limited from above
by the condition of applicability of the GL-expansion $|\tau|\ll 1$. There
exist a limitation from below. Even when criterion (12) is met average
fluctuation of the order parameter diverge when $|\tau|\to 0$ as
$1/\sqrt |\tau|$. As in conventional superconductors\cite{LarkOv}
fluctuations are small when
$|\tau|\gg \left(\lambda_{corr}^3/l_{tr}^2\xi_0\right)^2$.
Here $\lambda_{corr}$ is the correlation length for the random field
$\eta_{j l}({\bf r})$, $l_{tr}$ a transport mean free path for the Fermi
excitations. Assuming
 $\lambda_{corr}\sim$500\AA, $l_{tr}\sim$2000\AA, $\xi_0\sim$200\AA, one
 arrives at  inequality $|\tau|\gg$1/30. This estimation is
  not very accurate since the poorly known value of $\lambda_{corr}$
  enters it in the sixth power. More reliable  estimation comes from the
  experimentally observed smearing of the specific heat jump \cite{parp2},
  it renders $|\tau|\sim$1/25 for the lower limit.

The order parameter of ABM-phase Eq. (1) does not satisfy the criterion (12).
According to the above argument it can not be the order parameter of the
A-like phase. Possible forms of the order parameters suitable for a description
of  that phase were discussed elsewhere\cite{fom2,fom3}.
For the BW-order parameter criterion (12) is met. Identification of the B-like
phase as BW-phase agrees with the recent pulse-NMR experiments
 and with the observation of homogeneously precessing domain
(HPD) in that phase \cite{dmit}. Restriction imposed by criterion (12)
becomes essential
when B-phase is perturbed by magnetic field. Joint effect of aerogel and
 magnetic field on the BW-order parameter is discussed in the next section.

\section {B PHASE IN MAGNETIC FIELD}
In magnetic field two leading terms in the strength of the field have to be
added to the free energy Eq. (4):  linear in magnetic field and proportional to
imaginary part of the tensor $A_{\mu j}A_{\nu j}^*$:
$$
f_H^{(1)}=i\zeta e_{\mu\nu\lambda}A_{\mu j}A_{\nu j}^*H_{\lambda},  \eqno(14)
$$
and quadratic
in the field and proportional to real part of the same tensor
$$
f_H^{(2)}=\frac{1}{2}\kappa(A_{\mu j}A_{\nu j}^*+A_{\nu j}A_{\mu j}^*)
H_{\mu}H_{\nu}.           \eqno(15)
$$
Derivation of criterion (12) is based on the invariance of GL free energy with
respect to orbital rotations of the order parameter. The added terms
 preserve this property and as a consequence the criterion (12).
This criterion imposes restriction on possible changes of the
order parameter of B-phase. As in the case of pure $^3$He \cite{vollh} let
us search the order parameter in magnetic field in a form of B$_2$-phase
(or oblate state):
\begin{multline}
A_{\mu j}=\frac{e^{i\phi}}{2}\Bigl[
\Delta_{\uparrow\uparrow}(d_{\mu}+ie_{\mu})(m_j-in_j)+\\
\Delta_{\downarrow\downarrow}(d_{\mu}-ie_{\mu})(m_j+in_j)+
2\Delta_{\uparrow\downarrow}f_{\mu}l_j\Bigr]         \tag{16}
\end{multline}
Amplitudes
$\Delta_{\uparrow\uparrow}, \Delta_{\uparrow\downarrow},
\Delta_{\downarrow\downarrow}$
are found by minimization of the free energy.
The order parameter (16) satisfies criterion (12) if
$$
\Delta_{\uparrow\downarrow}^2=\frac{1}{2}\left(\Delta_{\uparrow\uparrow}^2+
\Delta_{\downarrow\downarrow}^2\right).                      \eqno(17)
$$
Separate analysis\cite{savin} shows that the ansatz (16) does not impose
additional restrictions on possible forms of the order parameter, provided a
deviation from the bulk B-phase is small.
Let us introduce following parametrization:
$\Delta_{\uparrow\uparrow}+\Delta_{\downarrow\downarrow}=2\Delta$,
$\Delta_{\uparrow\uparrow}-\Delta_{\downarrow\downarrow}=2\delta$,
then from eq. (17) $\Delta_{\uparrow\downarrow}^2=\Delta^2+\delta^2$.
Minimization of the free energy in a principal order on a small deviation
$\delta$ renders:
$$
\delta=\frac{-\zeta H}{2\Delta [3\beta_1+\beta_3-\beta_4+\beta_5]}. \eqno(18)
$$
Coefficient $\zeta$ is proportional to the derivative of the density of states
over energy, this makes $\delta$ really small. Keeping only principal order
correction we arrive for the order parameter to the expression:
$$
A_{\mu j}=\Delta R_{\mu j}+i\delta(e_{\mu}d_{\nu}-d_{\mu}e_{\nu})R_{\nu j}.
                                                                  \eqno(19)
$$
Eq. (19) shows that in a presence of aerogel magnetic field produces a change
of the BW-order parameter which is different from that in bulk $^3$He.  In the
bulk the order parameter remains real and principal effect is suppression of
$\Delta_{\uparrow\downarrow}$ pairing amplitude while
$\Delta_{\uparrow\uparrow}\approx\Delta_{\downarrow\downarrow}$.
In aerogel the increment of the order parameter is imaginary and principal
change is appearance of a difference $\delta$ between
$\Delta_{\uparrow\uparrow}$ and $\Delta_{\downarrow\downarrow}$.
In both cases the increments are small and hardly observable in a moderate
field, but there is an indirect manifestation of the induced change.
Rotation matrix $R_{\mu j}$ entering Eq. (2) is often parametrized by an angle
of rotation $\theta$ and direction of the rotation axis ${\bf n}$. The angle
$\theta$ is found by minimization of the dipole energy\cite{vollh} :
$$
f_D=\frac{3}{5}g_D(A_{\mu\mu}A_{jj}^*+A_{\mu j}A_{j\mu}^*).       \eqno(20)
$$
while direction of ${\bf n}$ remains arbitrary. This degeneracy is lifted
when B$_2$-order parameter (16) is substituted in the dipole energy (20).
For the order parameter (19) the relevant part of the dipole energy is
$$
f_{\bf n}=(9/2)g_D\delta^2 n_z^2.                                 \eqno(21)
$$
It has a minimum when $n_z=0$, i.e. when
${\bf n}\perp{\bf H}$ in a contrast to the pure $^3$He where the minimum
configuration is ${\pm\bf n}\parallel{\bf H}$. A strength of the orientational
effect is also different. Convenient characteristics of this strength is a
healing length $\xi_H$  i.e. a distance over which orientation of ${\bf n}$
recovers to its bulk value being perturbed by for example a boundary of a
sample. The healing length is found by comparison of the orientational energy
with the gradient energy. For the orientational energy in aerogel eq. (21) one
arrives at:
$\xi_H\sim\xi_0\frac{\tau}{\sqrt{5\lambda_D}\zeta H}$.
Dimensionless dipole constant (cf.\cite{vollh})
$\lambda_D\approx 5\cdot 10^{-7}$.  Coefficient $\zeta$ can be evaluated from
the splitting of A$_1$ -  A$_2$ transition in pure $^3$He
as 10$^{-3}$1/koe.  For H$\approx$1 koe and not too close to T$_c$ healing
length $\xi_H$ is of a few millimeters. This distance is comparable with
dimensions of samples used in most experiments. It means that orientation of
${\bf n}$ is determined by geometry of samples and not by the field.

\section {CONCLUSIONS}
Properties of superfluid phases of $^3$He in aerogel to a great extent are
determined by the spatial
fluctuations of the order parameter induced by aerogel.
This contrasts with the pure superfluid $^3$He where only average value of the
order parameter is important. In a presence of aerogel contribution of
fluctuations to physical quantities remains small only for a special class of
order parameters. Formal criterion for selection of such phases in a vicinity
of T$_c$ follows from very general assumptions about properties of aerogel. The
derived criterion rules out ABM-order parameter as a possible candidate for
A-like phase and imposes restriction on a possible variation of BW-order
parameter in magnetic field. Because of generality of the proposed scheme it
may be applied for description of the influence of a quenched disorder on
the other ordered systems with a continuous degeneracy.

\section {ACKNOWLEDGMENTS}
The author is grateful to V.V. Dmitriev, L.A. Melnikovsky and S.B. Savin for
useful discussions and comments.
This research was supported in part by RFBR grant (no. 01-02-16714)
 by Ministry of Education and Science of Russian Federation and by
 NATO grant SA(PST.CLG.979379)6993/FP.

\end{document}